\documentclass[12pt,preprint]{aastex}

\def\etal       {et al.}
\def\ie         {{i.e.},\ }
\def\eg         {{e.g.},\ }

\def\HII     {H~II}
\def\ammonia {NH$_3$}
\def\A       {$\alpha$}

\def\percc      {cm$^{-3}$}

\def\arcm{\ifmmode {' }\else $' $\fi}
\def\arcs{\ifmmode {'' }\else $'' $\fi}
\def\arcmper{\ifmmode \rlap.{'} \else $\rlap{.}' $\fi}
\def\arcsper{\ifmmode \rlap.{''} \else $\rlap{.}'' $\fi}
\def\porm   {\ifmmode\pm\else$\pm$\fi}
\def\kms    {\ifmmode{{\rm ~km~s}^{-1}}\else{~km~s$^{-1}$}\fi}
\def\masy   {\ifmmode{{\rm mas~y}^{-1}}\else{mas~y$^{-1}$}\fi}
\def\micron {\ifmmode{\mu{\rm m}}\else{$\mu$m}\fi}
\def\a      {\ifmmode {\rlap.}^{''}\! \else ${\rlap.}^{''}\!$\fi}

\newbox\grsign \setbox\grsign=\hbox{$>$} \newdimen\grdimen \grdimen=\ht\grsign
\newbox\laxbox \newbox\gaxbox
\setbox\gaxbox=\hbox{\raise.5ex\hbox{$>$}\llap
     {\lower.5ex\hbox{$\sim$}}}\ht1=\grdimen\dp1=0pt
\setbox\laxbox=\hbox{\raise.5ex\hbox{$<$}\llap
     {\lower.5ex\hbox{$\sim$}}}\ht2=\grdimen\dp2=0pt
\def\gax{\mathrel{\copy\gaxbox}}
\def\lax{\mathrel{\copy\laxbox}}

\slugcomment{2003 May 27}

\shorttitle{Dual Cometary \HII\ Regions}
\shortauthors{Cyganowski \etal}

\begin{document}


\title{Dual Cometary \HII\  Regions in DR21: Bow Shocks or Champagne Flows?}


\author{C.~J.~Cyganowski, M.~J.~Reid, V.~L.~Fish \& P.~T.~P.~Ho}
\affil{Harvard--Smithsonian Center for Astrophysics,
    60 Garden Street, Cambridge, MA 02138}
\email{cjc75@hermes.cam.ac.uk, reid@cfa.harvard.edu, vfish@cfa.harvard.edu,
       pho@cfa.harvard.edu}



\begin{abstract}

The DR~21 massive star forming region contains two cometary \HII\  regions,
aligned nearly perpendicular to each other on the sky.  This offers
a unique opportunity to discriminate among models of cometary \HII\  regions.
We present hydrogen recombination and ammonia line observations of
DR~21 made with the Very Large Array.
The velocity of the molecular gas, measured from \ammonia\ emission and
absorption lines, is constant to within $\pm1$~\kms\ across the region.
However, the radial velocity of the ionized material, measured from 
hydrogen recombination lines, differs by $\approx9$~\kms\ between the 
``heads'' of the two cometary \HII\  regions and by up to 
$\approx7$~\kms\  from that of the molecular gas. 
These findings indicate a supersonic velocity difference
between the compact heads of the cometary regions and between
each head and the ambient molecular material.  This suggests that the 
observed cometary morphologies are created largely by the motion of 
wind-blowing, ionizing stars through the molecular cloud, as in a bow 
shock model.

\end{abstract}



\section{Introduction}

Compact \HII\  regions are commonly observed in
massive star forming regions \citep{MST67,GL99}.  
\citet{RH85} first noted that
a compact \HII\  region in the G34.3$+$0.2 star forming region
had a cometary appearance, with a bright head and a diffuse tail.
Rather than being a unique or even a rare source, G34.3$+$0.2
became the archetypal example of a class of sources now
called cometary \HII\ regions.  The survey of \citet{WC89} revealed
that 20\% or more of compact \HII\  regions have cometary morphology.

Several models explaining the cometary appearance of compact \HII\ 
regions have been proposed and actively debated.  \citet{RH85} first
suggested that the cometary appearance of G34.3$+$0.2 could result from
relative motion between an ionizing star and its surrounding molecular 
material, analogous to elongated structures predicted for such stars moving
rapidly though the more diffuse interstellar medium \citep{WMC77}.  This
led to the ``bow shock'' model, which postulates supersonic motion
of a star, with a strong stellar wind, through a molecular 
cloud \citep{V90}.
An alternative model for the formation of cometary \HII\ regions
is the ``Champagne flow'' model.  This model assumes that the ionizing
star is nearly stationary with respect to the molecular cloud, but 
that the material surrounding the star has a steep density gradient 
\citep{I78,TT79}.
Since the size and shape of an \HII\  region are determined by the balance of 
ionization and recombination rates, an \HII\  region will be 
``extended'' in the direction of lowest density.  
In a uniform density gradient, this results in a
parabolic shape for the ionized region, which can appear cometary.
Other models for cometary \HII\ regions are discussed by 
\citet{GFC94,R86,GM87}.

%\clearpage
\begin{figure}
\epsscale{1.0}
\plotone{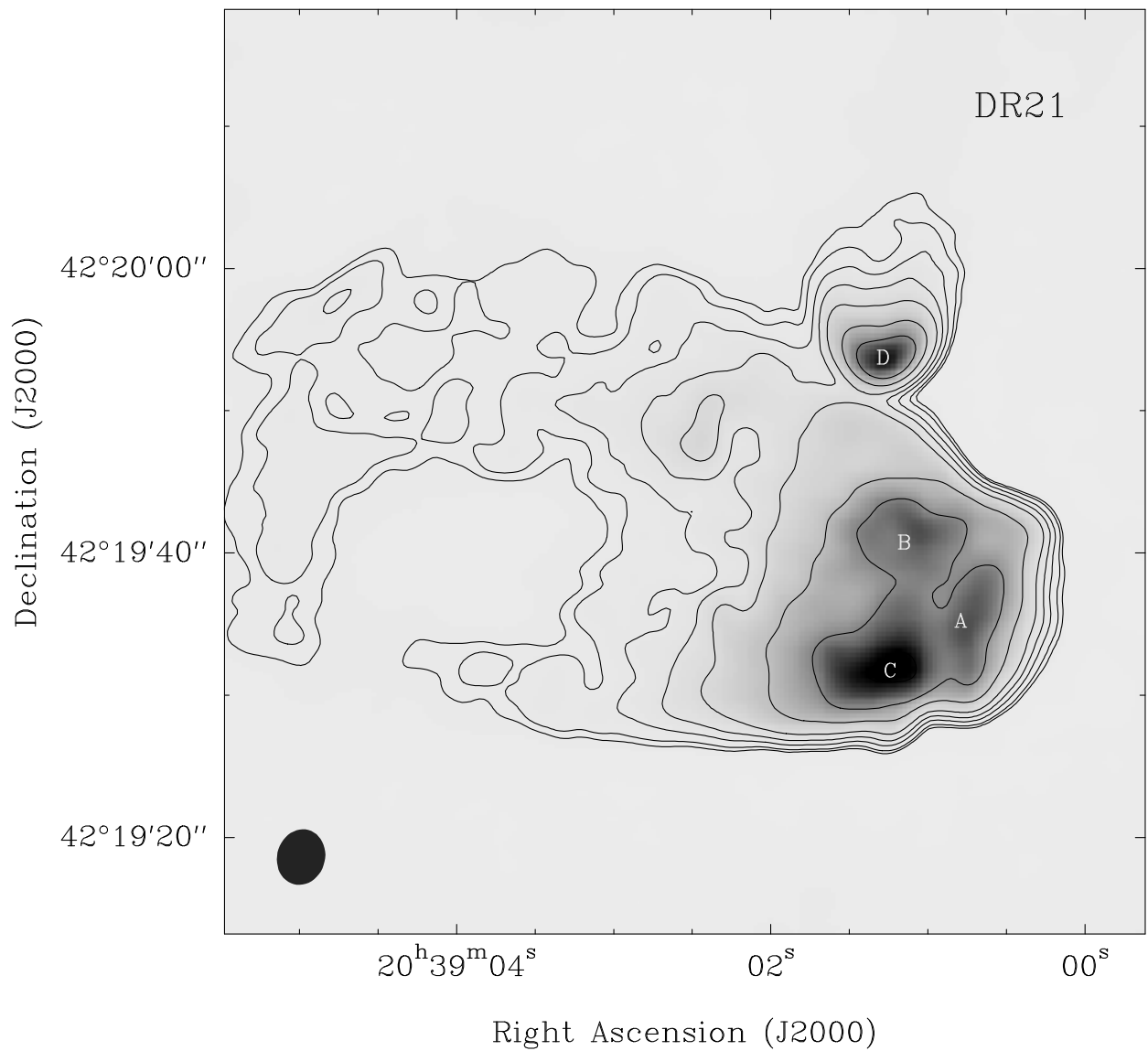}
\caption{Continuum emission from the dual cometary \HII\ regions
in DR~21 at 6-cm wavelength.  Contour levels are 0.01, 0.02, 0.05,
0.1, 0.2, 0.5, 1, and 2 Jy beam$^{-1}$.  The $\approx1\arcsper7$
FHWM beam is indicated by the filled ellipse near the lower left corner. 
Ionized condensations labeled A, B, C, \& D by Harris (1973) are 
indicated.
            \label{fig1}
        }
\end{figure}
  
\citet{H73} published a high resolution map at 5~GHz of the massive star
forming region DR~21.  She noted two \HII\ regions: a compact northern
(D) and an extended southern source, with the southern source resolved into 
``three very compact condensations'' (A, B, \& C).  Most discussion
in the literature of these \HII\ regions has followed the nomenclature
of Harris, as well as her interpretation of the southern \HII\ region's
condensations as being independent compact sources excited by
different stars.   We recently mapped this region with higher angular
resolution and sensitivity than Harris and discovered that the northern 
and southern \HII\ regions are both cometary \HII\ regions with nearly 
perpendicular symmetry axes projected on the sky. 
Fig.~1 shows the 6-cm wavelength continuum emission from these sources,
obtained from VLA B-configuration data taken on 2001 April 23.

Finding twin cometary \HII\ regions in one star forming region
provides a possibly unique opportunity to investigate the relative motion 
of cometary \HII\ regions, as well as their internal velocity structures 
and overall motion with respect to nearby molecular material.
Since the bow shock and Champagne flow models predict different velocity
structures, we undertook high angular resolution, spectral-line observations
in order to distinguish between these models.  We used the Very Large
Array (VLA) of the National Radio Astronomy Observatory\footnote{The
National Radio Astronomy Observatory is
a facility of the National Science Foundation
operated under cooperative agreement by Associated Universities, Inc.}
(NRAO) to map two high-frequency radio recombination lines, in order to
study motions in the plasma, and two ammonia (\ammonia) transitions, 
in order to study motions in the surrounding molecular material.
%The results of these observations favor the bow-shock interpretation as
%the primary shaping mechanism for these cometary \HII\ regions.

\section{Observations and Data Analysis}

We observed DR~21 in two hydrogen recombination lines and two
\ammonia\ transitions using the VLA on 2001 December 23.
The array was in the D-configuration, with a minimum baseline of
0.033 km and a maximum baseline of 1.03 km.  We obtained
approximately 1 hour of on-source integration time
each for the H53\A\/ line (42951.97 MHz rest frequency) and the H66\A\/
line (22364.17 MHz rest frequency).
The bandwidth for both recombination line observations was 12.5 MHz,
which was divided into 32 channels of 390.625 kHz (or 2.73 and
5.25~\kms\ for H53\A\ and H66\A, respectively).
The \ammonia\  (1,1) and (2,2) transitions 
were observed in both right and left circular polarization with bandwidths 
of 3.125~MHz, divided into 64 channels of 48.828~kHz (or 0.62~\kms),
and we obtained a total on-source integration time of $\approx2$ hours.

For both the recombination and \ammonia\  line observations,
the band center was set to an LSR velocity of $-2.0$ \kms\ .
The source 1331+305 was observed as a primary flux calibrator,
2015+371 as a secondary calibrator, and 2253+161 as a bandpass
calibrator. We used the NRAO Astronomical Image Processing System 
(AIPS) to edit, calibrate, image and display the data.
Initial calibration and subsequent self-calibration were done
using the ``channel 0'' data set (encompassing the central 75\% of the
original bandwidth).  

A preliminary flux scale for the H53\A\ observations at 43~GHz 
was based on a flux density of 1.47 Jy for 1331+305.  This flux density
is quite uncertain.  Therefore, after standard calibration, we measured the 
continuum flux density of the brightest portion of the southern
cometary \HII\ region at 22 and 43~GHz and adjusted the 43~GHz
fluxes to give the expected optically thin spectral index of $-0.1$
for an \HII\ region.  This required multiplying the H53\A\ 
data by a factor of 1.24.

Line-minus-continuum ({\it u,v})-databases were created by fitting a straight
line to the (continuum) signal in spectral channels with little or no
line emission or absorption and then subtracting this fit from
the entire spectrum.  Because the recombination lines are very
broad, we could only use a few channels on each side of the H66\A\ 
line to measure the continuum.  We used the same observing bandwidth 
for the H53\A\ observations and were able to use only a few channels 
near the low-frequency end of the spectrum as off-line channels.  Thus,
we removed only a constant continuum level for the H53\A\ data.  
(However, as discussed below, for both recombination
lines we ultimately removed any residual continuum emission by
fitting a sloped baseline to spectra generated from the 
line-minus-continuum maps.)  A pure continuum ({\it u,v})-database was also
constructed from the off-line channel baseline fits, evaluated at the 
center frequency of the original bands.  A 22~GHz continuum map made from 
the baseline fits for the H66\A\ data is shown in Fig.~2.

%\clearpage
\begin{figure}
\epsscale{1.0}
\plotone{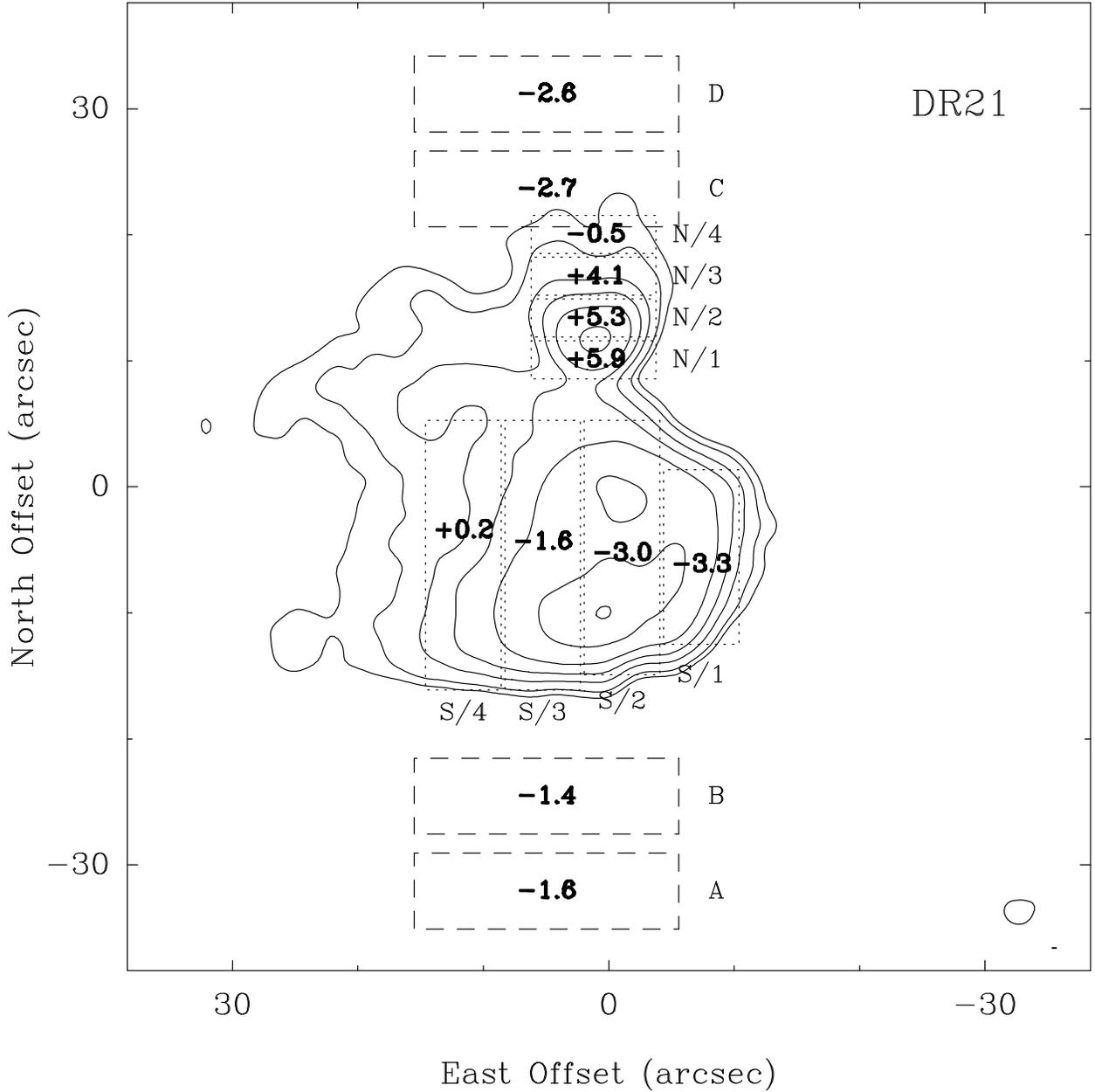}
\caption{Continuum emission from the dual cometary \HII\ regions
in DR~21 at 1.3-cm wavelength.  Contour levels are 0.02, 0.05, 0.1,
0.2, 0.5, 1, and 2 Jy.  Dotted boxes indicate ``long-slit'' areas
over which spectra were summed for Gaussian fitting of the hydrogen 
recombination lines.   Dashed boxes indicate the areas
where \ammonia\ emission spectra were generated; areas for \ammonia\ 
absorption spectra cover most of the northern and southern \HII\ regions
and correspond approximately to regions N/1 to N/3 and S/1 to S/3,
respectively.
Labels near the boxes indicate position listed in Table 1.
Line-center velocities from Gaussian fitting are given in \kms\ with
respect to the LSR in bold face in the center of the boxes.  For the
hydrogen recombination velocities, the average of H66\A\ and H53\A\ 
fits are indicated, where both lines were detected.  
            \label{fig2}
        }
\end{figure}
  
The calibrated line-minus-continuum data were used to generate image cubes.
For the hydrogen recombination line data,
we made image cubes with tapered (6\arcs\ beam) and un-tapered 
(3\arcs\ beam) ({\it u,v})-data, and used the tapered cubes to
analyze the large southern \HII\ region and the un-tapered cubes for the
smaller northern \HII\ region. Next, we generated spectra by summing 
the emission over ``long-slits,'' approximately perpendicular to the 
major axes of the two cometary \HII\ regions.
We chose such long-slits in order to increase signal-to-noise ratios
and to facilitate comparison to models which generally have
axial symmetry about the cometary axis.  
The positions of these long-slits are indicated on Figure~2 and the
resulting spectra for the H66\A\ transition are shown in Figure~3. 
In order to determine Doppler velocities, we fitted a Gaussian line
profile to these spectra.
The amplitude, central velocity, and FWHM of the Gaussian 
line profile were adjustable parameters, 
as well as two parameters allowing for a linear spectral baseline.
For both recombination lines, the spectral baselines typically sloped by about 
$\pm10$\% of the peak amplitude of the line.  
Results of the Gaussian fits are given in Table~1 and Figure~3.
While formal fitting errors usually
were $<0.3$~\kms, we estimate a more realistic error that allows
for modeling uncertainty of $\approx1$~\kms.

We analyzed both \ammonia\  transitions and obtained nearly identical
kinematic information from the two transitions, and for brevity we 
report here only the \ammonia\ (1,1) transition results.
The \ammonia\ (1,1) transition main-hyperfine component at a rest frequency 
of 23694.496 MHz was used to obtain Doppler velocities.  
Offset from the continuum sources, we 
detected the \ammonia\ in emission and produced spectra
summed over long-slits (see Fig.~2), as for the
hydrogen recombination lines.  Toward the cometary
\HII\ regions, we detected \ammonia\ in absorption and spectra were 
obtained by summing over regions that enclosed most of the continuum 
emission from each region.  The spectra showed a small, but
distinct, baseline curvature.  Thus, we fitted a second-order baseline
in addition to Gaussian lines to the spectra.
The Gaussian fits to the main-lines are given in Table~1.
In Figure~3 we show the observed spectra and the best-fitting (three) Gaussians
for the \ammonia\ main- and inner satellite-hyperfine components toward 
the \HII\ regions and the H66\A\ lines at positions along the cometary axes 
of the southern and northern \HII\ regions.

%\clearpage
\begin{figure}
\epsscale{1.0}
\plotone{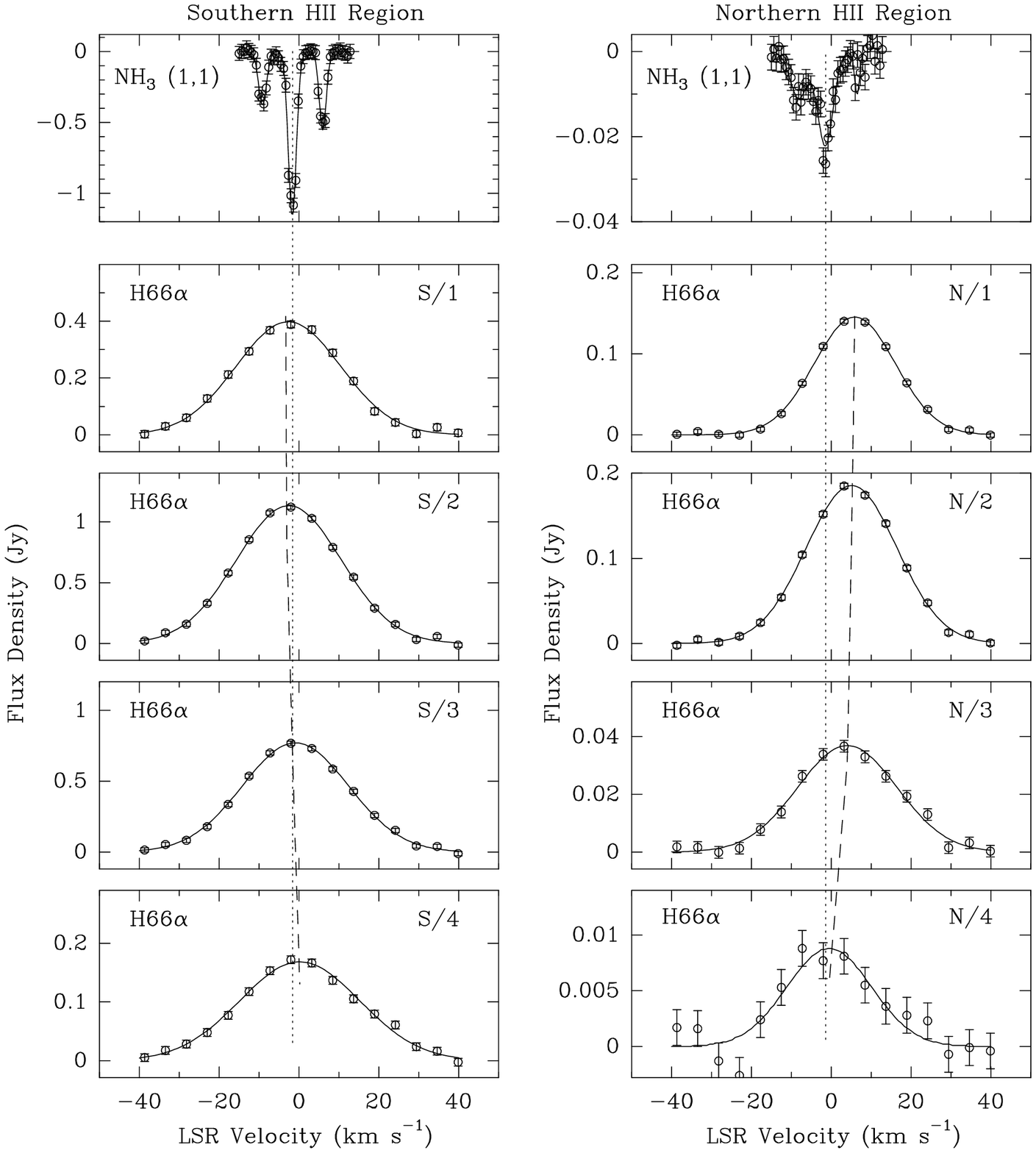}
\caption{\ammonia\ and H66\A\  spectra toward the southern ({\it left panels})
and northern ({\it right panels}) cometary \HII\ regions in DR~21.
Dotted vertical lines give the velocity for the molecular material 
as indicated by the \ammonia\  main-hyperfine line absorption.
Dashed sloping lines indicate the velocity of the ionized material,
from the average of the H66\A\ and H53\A\ line fits, 
as one observes from head toward tail (from top to bottom plots) 
along the cometary axes of the \HII\ regions.
            \label{fig3}
        }
\end{figure}

\begin{deluxetable}{lccc}
\tablecolumns{4}
\tablewidth {4truein}
%\rotate
\tabletypesize{\scriptsize}
%\tablenum{1}
\tablecaption{Gaussian Fit Parameters$^1$}
\tablehead{
%%\multicolumn{4}{c}{H$66\alpha$} \\
%%\cline{1-4} \\
\colhead{HII Region/} &\colhead{S} &\colhead{v$_{LSR}$} &\colhead{FWHM} \\
\colhead{Position}       &\colhead{(Jy)}      &\colhead{(\kms)} &\colhead{(\kms})
}
\startdata
\cutinhead{H$66\alpha$}\\
S/1     & $0.398\pm0.009$  & $-2.6\pm0.3$  & $30.9\pm1.0$ \\
S/2     & $1.132\pm0.013$  & $-2.5\pm0.2$  & $30.9\pm0.5$ \\
S/3     & $0.770\pm0.011$  & $-0.8\pm0.2$  & $31.3\pm0.7$ \\
S/4     & $0.168\pm0.005$  & $+0.2\pm0.5$  & $35.4\pm1.6$ \\
\\
N/1     & $0.145\pm0.002$  & $+5.9\pm0.2$  & $23.9\pm0.4$ \\
N/2     & $0.186\pm0.002$  & $+5.1\pm0.2$  & $26.7\pm0.5$ \\
N/3     & $0.038\pm0.001$  & $+4.1\pm0.6$  & $29.3\pm1.6$ \\
N/4     & $0.009\pm0.001$  & $-0.5\pm1.6$  & $29.3\pm4.5$ \\
\cutinhead{H$53\alpha$}\\
%   Flux densities scaled by 1.24 (see text)
S/1     & $0.686\pm0.027$  & $-3.9\pm0.4$ & $33.4\pm1.6$ \\
S/2     & $2.075\pm0.029$  & $-3.6\pm0.2$ & $31.9\pm0.6$ \\
S/3     & $1.266\pm0.016$  & $-2.2\pm0.2$ & $31.5\pm0.6$ \\
\\
N/1     & $0.219\pm0.004$  & $+5.8\pm0.2$  & $23.2\pm0.5$ \\
N/2     & $0.301\pm0.006$  & $+5.5\pm0.2$  & $26.3\pm0.7$ \\
\cutinhead{NH$_3$ (1,1)}
A               & $+0.116\pm0.004$ & $-1.6\pm0.1$ & $1.3\pm0.1$ \\
B               & $+0.188\pm0.008$ & $-1.4\pm0.1$ & $1.3\pm0.1$ \\
Southern        & $-1.157\pm0.038$ & $-1.6\pm0.1$ & $2.4\pm0.1$ \\
Northern        & $-0.022\pm0.002$ & $-1.4\pm0.2$ & $4.5\pm0.6$ \\
C               & $+0.079\pm0.005$ & $-2.7\pm0.1$ & $1.4\pm0.1$ \\
D               & $+0.023\pm0.004$ & $-2.6\pm0.2$ & $2.7\pm0.6$ \\
\\
\tablenotetext{1}{Fits to spectra obtained by summing the spectral 
brightness over the regions indicated graphically
in Fig.~2. \ammonia\ spectra at positions labeled ``Southern''
and ``Northern'' were obtained by summing the spectral brightness 
over most of the detected \HII\ emission of the southern and
northern \HII\ regions, respectively.
}
\enddata
\end{deluxetable}

\section{Discussion}

The first explanation proposed for the cometary \HII\  region morphology
\citep{RH85}  invoked relative
motion between an ionizing star and its external molecular environment.
For the archetypal cometary \HII\ region, G34.3$+$0.2, \citet{RH85} 
noted the presence of a supernova 
remnant to the west of the cometary region's ``head'' and suggested that
a stellar wind from the supernova's precursor star might be responsible
for the \HII\  region's cometary shape.  The bow shock model,
developed to explain cometary \HII\ regions \citep{V90,ML91,VM92},
expanded upon this suggestion.  It requires highly supersonic 
motion of a wind-blowing, ionizing star through dense surrounding material.  
On the other hand, a cometary appearance
can also result from the ionization/recombination balance in a region
with a strong density gradient, often called a Champagne flow 
\citep{I78,TT79}.  In addition, other forces may play a role in
shaping cometary \HII\ regions \citep{GFC94}.

In this section, we first address the possibility that, in regions
where ionized gas flows rapidly, hydrogen recombination line
velocities may be difficult to interpret.  Next,
we briefly outline the dominant kinematic features of the
bow shock and Champagne flow models.  Finally, we compare the observed 
velocities of the ionized and molecular material to these models
and argue that they are mostly consistent with the predictions of
the bow shock model.  

\subsection{Are Recombination Lines Reliable Velocity Indicators?}

We observed two hydrogen recombination lines in order to assess 
if the velocities measured by these lines differ significantly.
\citet{BE83} and \citet{K95} discussed
such an effect for W3OH, with line-center velocities shifting 
by $\approx22$~\kms\ between the H110\A\ and H35\A\ lines.
The shift in observed line-center velocity with transition is 
large for high principal quantum number (low frequency) 
transitions and asymptotically approaches a constant at low
principal quantum number (high frequency) transitions.  
Keto et al.~suggest that this transition-dependent velocity 
shift results from a combination of a large velocity gradient 
in the ionized material and a strong increase in line-width 
for higher principal quantum number transitions, owing to collisional 
broadening.   Since, for electron densities $\lax10^7$~\percc,
collisional broadening is small for the H66\A\ and lower principal
quantum number transitions, we would expect that our recombination lines 
would not be sensitive to this complication.
%For example, in W3OH, \citet{GRM85} find only a
%difference of $\approx2$~\kms\ between the velocities derived from
%H76\A\/ and H66\A\/ observations.  In general, both theory and observations 
%indicate that hydrogen recombination line velocities become less 
%problematic for lower principal quantum numbers 
%(or higher line frequencies).

For DR~21, we observe almost no shift in velocity between the
H66\A\/ and H53\A\/ lines.  The velocities of the H66\A\/ and
H53\A\/ lines agree within 2~\kms\  at all locations.   
For the northern \HII\ region, the agreement is within the
joint uncertainties ($<0.3$~\kms).
For the southern \HII\ region, the H53\A\  lines appear blue 
shifted by about $1.3$~\kms\  compared to the H66\A\ lines at 
the same positions on the sky.
The spectral baselines in the two transitions tended
to slope in opposite directions, and small correlations between the
spectral slopes and the center velocities might contribute to these
small shifts.  

\citet{RGG89} mapped H76\A\ recombination lines toward several positions
over the DR21 \HII\ regions.   The dual-cometary structure, clearly
visible in our Fig.~1 at 6~cm wavelength, is not as evident in their
14.7~GHz continuum image, and they did not use long-slits 
to measure velocities along the symmetry axes of the cometary \HII\
regions.  Instead they formed spectra at local peaks in the radio brightness,
including one in the northern and five in the southern \HII\ region.
Toward the northern \HII\ region (their position D) the H76\A\ line 
peaked at $4.6$~\kms, in good agreement with the average of our regions 
N/1 through N/3 which span velocities of 4.1 to 5.9~\kms.  
Toward the head of the southern \HII\ region, \citet{RGG89} 
measured a velocity (at their position A) of $-4.1$~\kms\ which 
compares well with our measurement of $-3.3$~\kms\ for region S/1.  
Also, their H76\A\ measurements at positions B and C 
yielded velocities of $-1.2$ and $-3.2$~\kms; these positions if combined
correspond to summing our regions S/2 and S/3 for which we measure 
velocities of $-1.6$ and $-3.0$~\kms.  Again this indicates good
agreement between the H76\A\ measurements of \citet{RGG89} and
ours at H66\A\ and H53\A.

In conclusion, since (1) on average the H76\A, H66\A, and H53\A\ 
lines are in agreement to within $\lax1$~\kms, (2) the widths 
of these lines are nearly identical, and (3)
exceptionally high densities would be required to yield any 
substantial velocity shift between H66\A\/ and H53\A\/ lines,
we conclude that the velocities measured by these high-frequency
recombination lines are not significantly affected by strong velocity 
gradients coupled with collisional broadening.   
Thus, the lines should indicate the line-of-sight velocity of the 
ionized material in the dense \HII\ regions to within about $1$~\kms.

\subsection{The Bow Shock Model}

The bow shock model, as formulated by \citet{V90}, 
postulates a balance between the stellar wind pressure and the
ram pressure of dense molecular material caused by the motion 
of the star through a dense molecular cloud. This
gives rise to an \HII\  region in the form of a thin, paraboloidal shell.
The volume between the thin ionized shell and the star
is mostly evacuated by the stellar wind, and this model predicts  
limb-brightening in the tail, which is characteristic of cometary \HII\ regions
(cf. Fig.~1).

%In order to produced a reasonably collimated tail, the motion
%of the star with respect to the ambient cloud must be comparable to,
%or exceed, the speed at which the \HII\ region tends to expand
%(in the absence of relative motion).
%If the \HII\ region expands freely at its internal sound speed,
%owing to its high internal pressure, then
%the star must move faster than $\approx10$~\kms\ relative to
%its surroundings.  However, if other forces impeed expansion,
%such as gravity close to the ionizing star \citep{K02} or magnetic pressure 
%(\eg\ RMB88), perhaps one could observe a cometary \HII\ region morphology 
%with a somewhat lower stellar speed relative to the molecular cloud.

%\clearpage
\begin{figure}
\epsscale{1.0}
\plotone{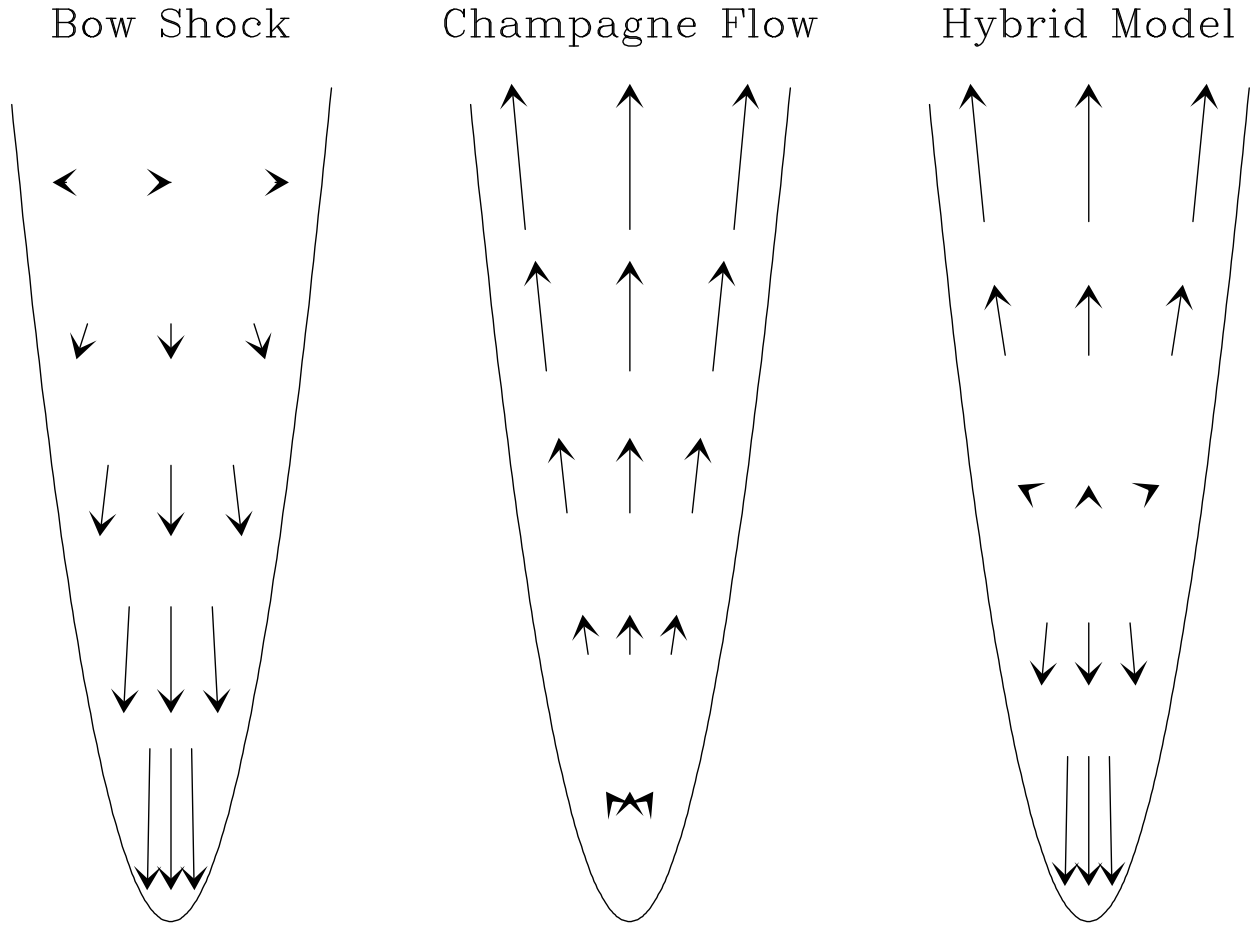}
\caption{Schematic representations of the velocity field in cometary
\HII\ regions for three models: a bow shock ({\it left}), a
Champagne flow ({\it center}), and a hybrid model ({\it right}).
The velocity reference for all models is the ambient molecular
gas, presumed to surround the cometary \HII\ regions.
            \label{fig4}
        }
\end{figure}
  
The bow shock model predicts that the ionized material near the
cometary head should be moving, on average, with the velocity of the
star (and hence the ionized gas should be moving rapidly with respect 
to the molecular gas of the surrounding cloud).  
Further down the cometary tail the velocity of the ionized gas
should approach that of the molecular cloud (see Fig.~4).

\subsection{The Champagne Flow Model}

The Champagne flow model evolved from ``blister'' models for
compact \HII\ regions \citep{I78} and also can yield 
cometary morphology \citep{TT79,YTB83}.  
The Champagne flow model postulates that cometary \HII\ regions 
form when ionized gas expands asymmetrically out of a dense clump 
within a molecular cloud into a lower-density region.
In this model, the velocity of ionized gas near the cometary head 
should be close to the velocity of the densest molecular gas, 
whereas down the cometary tail, the ionized gas should attain
high velocities of $\approx30$~\kms\ as the ionized material is 
accelerated by a strong pressure gradient through the ``nozzle'' 
\citep{BTY79,YTB83}.
Thus, the predictions of the Champagne flow model, 
shown schematically in Fig.~4, are nearly opposite those of 
the bow shock model with respect to the velocity field.

\subsection{The DR~21 Cometary \HII\ Regions}

The Doppler velocities of the \ammonia\ lines seen in absorption
toward the cometary \HII\ regions are $\approx -1.5$~\kms,
and the emission lines north and south of these \HII\ regions
differ by $\lax 1$~\kms\ from this value.  Thus, the molecular
material, presumably in close proximity to the ionized material,
shows little velocity structure.  On the other hand, the ionized
material, as indicated by hydrogen recombination lines, 
is clearly kinematically different from the molecular material.
The northern \HII\ region has a velocity, at the brightest emission
near the cometary head, which is red shifted by 7~\kms\ with respect
to the molecular gas, whereas the southern \HII\ region appears blue 
shifted by 2~\kms.  Assuming random orientations for the 
velocity vector of these \HII\ regions (with an average inclination
to the line of sight of $60^\circ$), this suggests space velocities
(3-dimensional speeds) of about 15 and 4~\kms\ relative to the 
molecular material.  This supports a bow-shock model and opposes a 
Champagne flow model.

Since the cometary appearance of an \HII\ region becomes more
pronounced as the structure is viewed from the side
(\ie the closer the cometary axis is to being in the plane of the sky), 
the space velocities are likely to be greater than those estimated for
random orientations.  Indeed, the long cometary ``tail'' of the
southern \HII\ region, compared to the short ``tail'' of the
northern \HII\ region, suggests that the southern \HII\ region
may have its cometary structure nearly in the plane of the sky.
If this is true, then most of its space velocity may be in 
the plane of the sky, and only a small component of the
space velocity would be indicated by its Doppler velocity.
Thus, it would not be surprising if the space velocity of the
southern \HII\ region is $\sim5\rightarrow10$~\kms\ with respect to the
ambient molecular material, as for the northern \HII\ region.

For both \HII\ regions, the measured velocities change 
with position along the cometary axis.  
Starting at the head and moving down the tail
of the cometary \HII\ regions, the velocities approach that 
of the ambient molecular material.  This is in keeping with
the bow shock model.  For the southern \HII\ region, there
is an indication, at position S/4, that the velocity of the ionized material
``drifts past'' that of the ambient molecular material, as indicated
by the \ammonia\ velocities.  \citet{RGG89} report H76\A\ lines
from two positions further down the tail than our S/4 position
with velocities of $12.6$ and $4.2$~\kms.  
This could indicate the beginning
of a pressure driven outflow down the tail of the cometary \HII\ region,
as suggested by Champagne flow models.  Indeed, there is no reason
that a bow shock and Champagne flow could not operate in the same
cometary \HII\ region (\eg as in the hybrid-model in Fig.~4).  
However, more sensitive hydrogen recombination line observations 
than we have achieved will be needed to measure velocities of the ionized
material further down the cometary tails.

\section{Complications and Future Work}

In this paper, we have analyzed the velocity of the ionized material 
only along the cometary axes of the \HII\ regions in DR21.
For the northern cometary \HII\ region, there is no
evidence for velocity structure perpendicular to the cometary
axis.  However, for the southern cometary \HII\ region, there
are indications of velocity structure perpendicular to the cometary
axis.  In this source, deviations from axial symmetry seem to 
grow in the weak, diffuse tail.  Given our current sensitivity, 
we cannot properly characterize the kinematic asymmetry of the 
ionized gas in the cometary tails.  

Deviations from axial symmetry are also seen in the archetypal 
cometary \HII\ region G$34.2+0.2$ \citep{GRG86}.
Neither the bow shock nor the Champagne flow models, at least
in their simplest forms, predict such structure.  
However, one might expect complex kinematics for the ionized material 
in cometary \HII\ regions, owing to possible anisotropic stellar
winds, the effects of magnetic fields, and/or mis-oriented 
stellar motions and molecular cloud density gradients.  
Perhaps future observations with the VLA, which can yield
a factor of $\gax3$ increase in sensitivity over our current
data, will better characterize these effects and lead to a more 
complete understanding of cometary \HII\ regions.

The widths of hydrogen recombination lines may provide information 
which can help discriminate among various models.  Thermal line-widths 
for an 8000~K hydrogen plasma are expected to be $\approx20$~\kms.  
We observed broader line-widths of $\approx30$~\kms\ for both cometary HII 
regions, indicating non-thermal components of $\approx20$~\kms.   
This could be caused by the effects of flows within the HII regions,
driven by strong stellar winds.  Note that the measured line-widths are
relatively constant over the regions measured.  The bow shock model, 
in its simplest form, predicts line-widths greatest at the head
and decreasing toward the tail of a cometary \HII\ region.   
Alternatively, the Champagne flow model predicts line-widths 
greatest down the tail as material is strongly accelerated.  
Thus, the line-width data do not strongly support either model.
However, as discussed above, plasma flows within the cometary \HII\ 
regions may be quite complex and will complicate interpretation of 
line-widths.

Finally, on angular scales $\approx10$ times larger 
than considered in this paper, the DR21 star forming region displays 
rich and complex structure.  In a series of papers, Garden and collaborators
\citep{G-I,G-II,G-III,G-IV} discovered that the DR21 \HII\ regions
are seen projected toward the center of a structure about 6\arcmin\
long oriented approximately in the NE--SW direction.  
This structure was detected in vibrational H$_2$ as well as
rotational CO and HCO$^+$ line emission.  They argue that this structure is
an extremely luminous bipolar outflow associated with the southern \HII\ 
region discussed in this paper.  Given the location of the southern \HII\
region, and the close alignment of its cometary axis with that of the
axis of elongated structure, it is reasonable to consider the star
(or stars) powering the \HII\ region as responsible for a bipolar
outflow that excites the  vibrational H$_2$ emission.
However, in this case it is difficult to reconcile the sharp western edge 
of the southern cometary \HII\ region with the implied bipolar
outflow.  Unless the western outflow was formed prior to the \HII\ region,
the outflow would need to pierce the edge of the \HII\ region 
without disrupting it or producing an observable effect.  
This seems unlikely, but needs further analysis.

\citet{G-II} point out that the vibrational H$_2$ emissions of
DR21 and Orion are similar.  For the case of Orion, even though
the source is much closer than DR21, the origin of the high
velocity flow seen both in H$_2$O masers and vibrational H$_2$ emission
is unclear \citep{GRM81,NNG82}.  Moved to DR21's distance, some
of the candidate sources for the outflows in Orion, 
such as source--I and source--N \citep{MR95},
would be separated by less than $\approx1$\arcsec.  
Neither of these sources excites a strong \HII\ region.  
Were a different source to excite a strong \HII\ region in Orion, 
similar to the southern cometary
\HII\ region in DR21, the candidate sources would be difficult
to distinguish, let alone identify.  Thus, one possible explanation
for the approximate coincidence of the southern \HII\ region with the 
center of the elongated bipolar-like structure in DR21 is that they are 
excited by different sources at different depths in the same massive 
star forming region and are otherwise unrelated.

\acknowledgments

\end{document}